\documentclass[12pt]{iopart}

\usepackage{iopams}
\usepackage{graphicx}

\newcommand{\beq}{\begin{equation}}
\newcommand{\eeq}{\end{equation}}
\newcommand {\ba} {\begin{eqnarray}}
\newcommand {\ea} {\end{eqnarray}}

\begin{document}

\title[Optical properties of LiFeAs
single crystal]{Optical properties of the iron-based
superconductor LiFeAs single crystal}

\author{Byeong Hun Min$^{1}$,
Jong Beom Hong$^{2}$,Jae Hyun Yun$^{3}$,Takuya
Iizuka$^{4}$,Shin-ichi Kimura$^{4,5}$,Yunkyu
Bang$^{3,6,\dag}$,Yong Seung Kwon$^{1,\S}$}

\address{$^{1}$Department of Emerging Materials Science,
Daegu Gyeongbuk Institute of Science and Technology (DGIST), Daegu 711-873, Republic of Korea\\
$^{2}$Department of Physics, Sungkyunkwan University, Suwon 440-746, Republic of Korea\\
$^{3}$Department of Physics, Chonnam National University, Kwangju 500-757, Republic of Korea\\
$^{4}$ School of Physical Sciences, The Graduate University for Advanced Studies, Okazaki 444-8585, Japan \\
$^{5}$ UVSOR Facility, Institute for Molecular Science, Okazaki
444-8585,Japan\\ $^{6}$ Asia Pacific Center for Theoretical
Physics, Pohang 790-784, Republic of Korea}
\ead{$^{\dag}$ykbang@chonnam.ac.kr}
\ead{$^{\S}$yskwon@dgist.ac.kr}

\begin{abstract}
We have measured the reflectivity spectra of the LiFeAs ($T_c$ =
17.6 K) single crystal in the temperature range from 4 to 300 K.
In the superconducting (SC) state ($T < T_c$), the clean opening
of the optical absorption gap was observed below 25 cm$^{-1}$,
indicating an isotropic full gap formation. In the normal state
($T > T_c$), the optical conductivity spectra display a typical
metallic behavior with the Drude type spectra at low frequencies,
but we found that the introduction of the two Drude components
best fits the data, indicating the multiband nature of this
compound. A theoretical analysis of the low temperature data ($T$
= 4 K $< T_c$) also suggests that two SC gaps best fit the data
and their values were estimated as $\Delta_1$ = 3.3 meV and
$\Delta_2$ = 1.59 meV, respectively. Using the
Ferrell-Glover-Tinkham (FGT) sum rule and dielectric function
$\epsilon_{1}(\omega)$, the plasma frequency of the SC condensate
($\omega_{ps}$) is consistently estimated to be 6,665 cm$^{-1}$,
implying that about 65 \% of the free carriers of the normal state
condenses into the SC condensate. To investigate the various
interband transition processes (for $\omega >$ 200 cm$^{-1}$), we
have also performed the local-density approximation (LDA) band
calculation and calculated the optical spectra of the interband
transitions. This theoretical result provided a qualitative
agreement with the experimental data below 4000 cm$^{-1}$ .
\end{abstract}

\pacs{74.25.Gz,74.20.Rp,74.70.Xa}
\submitto{\NJP}
\maketitle

\section{\label{sec:intro}Introduction}

Iron based superconductors, i.e. ReFeAsO$_{1-x}$F$_x$ (Re:
rare-earth elements), AFe$_2$As$_2$ (A: alkali metal), FeSe, and
AFeAs (A: alkali metal)
\cite{kamihara,ren,chen,rotter,ren2,leeh,hsu,song1,tapp}, have
become the focus of intensive research in the hope of
understanding the pairing mechanisms of the unconventional
superconductivity based on the 3$d$ electrons. Among the iron
based superconductors, LiFeAs is unique because of its simple
crystalline structure and its moderately high superconducting (SC)
transition temperature ($T_c \sim$  17 K) without doping. The
theoretical band calculations have predicted a multiband nature
(up to a maximum of five conduction bands) of this compound, which
was confirmed experimentally \cite{singh1,nakamura,borisenko}.
Hence the various SC properties are expected to show the
multi-bands and multi-gaps nature such as: specific heat
measurements ($C_p$), tunnel diode resonator measurement (TDR),
microwave surface impedance measurements, lower critical field
studies ($H_{c1}$), Raman spectroscopy and angle resolved
photoemission spectroscopy (ARPES)
\cite{borisenko,wei,dong2,imai,kimh,sasmal,song2,um}. However, in
most of experimental measurements, except ARPES, the SC gaps were
estimated indirectly, and therefore the estimated values should
contain a degree of uncertainty. Even in the case of ARPES,
although the SC gap sizes are directly measured, it probes only
surface states and also has the problem of the surface
degradation.

Far infrared spectroscopy is a powerful tool for investigating the
bulk properties of the electronic structure of materials such as
the changes of the Fermi surfaces (FSs) and low energy
excitations, and hence it can directly measure the SC energy gap
sizes of the bulk state. The studies of the SC gap by the far
infrared spectroscopy have already been reported for 1111, 122, 11
and 245 systems
\cite{qazilbash,dong,barisic,kimk,wu,li,tu,homes,maksimov,Charnukha,Chen}
but until now it has not yet been reported for LiFeAs system. In
the case of LiFeAs, especially, the accurate measurement of the
spectrum below 100 cm$^{-1}$ is very important for examining the
SC gap(s) and the multiband nature because its $T_c$ is relatively
lower than other iron based superconductors studied with the
optical spectroscopy.

In this study, we report the optical measurements of LiFeAs single
crystal with $T_c$ = 17.6 K. In particular, we have measured the
reflectivity data down to 15 cm$^{-1}$ in order to resolve the SC
gap formation. To control the uncertainty level at far-infrared
and tera-hertz frequency region, we have used our specially
designed feedback positioning system \cite{kwon}. Below $T_c$, we
have observed the clear signatures of the SC gap formation in the
reflectivity data which becomes flat and approaches unity at low
frequencies. In the real part of the optical conductivity, which
is obtained by the Kramers-Kronig (KK) relation from our
reflectivity data, this feature is identified as the opening of
the optical absorption gap at low frequencies. Theoretical fitting
using the generalized Mattis-Bardeen formula \cite{zimmermann}
yields the best result with two SC gaps with the estimated sizes
as 1.59 and 3.3meV, respectively. These two gap values are in
perfect agreement with the estimate from the specific heat
measurement \cite{Stockert}. Also the larger gap value is
consistent with the already reported results
\cite{borisenko,wei,imai,sasmal,song2,kimh,inosov} by other
experimental probes. We also showed that the normal state optical
conductivity can be best understood by introducing two Drude
spectral components, which consistently supports the multiband
nature of LiFeAs superconductor.

\section{Experimental details}

The single crystal of LiFeAs was grown by a closed Bridgeman
method \cite{song1} and the size of obtained single crystal is
approximately 3 mm $\times$ 3 mm with shiny surface. The single
crystal has a layer structure with a cleaved surface perpendicular
to $c$-axis, [001] direction. Performing the optical measurement
on LiFeAs is a challenge due to its quick degradation in air
\cite{kimh}. In order to avoid the degradation of the sample, it
was cleaved in a high purity Helium gas filled glove bag and the
sample was attached to the optical sample holder under Helium gas
atmosphere. Electrical resistivity measurement was carried out
using the standard four probe method. Reflectivity measurements
were carried out on the freshly cleaved single crystal surface
(${ab}$-plane). JASCO FTIR610 was used for the infrared
reflectivity spectroscopy in the frequency range from 40 to 12,000
cm$^{-1}$ for temperatures of 4 $\sim$ 300 K. To improve accuracy
in the frequency region below 100 cm$^{-1}$, JASCO FARIS was used
for the THz frequency range from 15 to 200 cm$^{-1}$ for the same
temperatures. ACTON VM 504 spectrometer was used for visible and
violet reflectivity spectroscopy in the frequency range from 9,000
to 40,000 cm$^{-1}$. An in-situ gold evaporation technique was
used to calibrate for the absolute reflectivity value. Sample size
of most of the iron based superconductors, including ours, is
smaller than the beam size ($\phi \sim$ 8 mm), so that the
interference from the sample edge and the on-passing optical
window occurs and becomes the main source of uncertainty. Thus we
have specially designed the feedback positioning system to
suppress this type of uncertainty \cite{kwon,kimura}. A small
reference mirror, which will be used for locating the reference
point, is attached on the opposite side of the sample holder and
thus searches for the maximum intensity of the reflected laser
with Si diode. The resolution of stepping motor being used for
vertically shift is 0.1 $\mu$m/step. Once the exact distance
between reference mirror and sample is known, we can always find
the same vertical position of the sample. Using this feedback
method, we could reduce the uncertainty level down to 0.6 \% and
0.3 \% below and above 100 cm$^{-1}$, respectively.

\section{Results and Discussions}
\subsection{Reflectivity and Optical Conductivity}

Figure 1 shows the reflectivity spectra, $R(\omega)$, of LiFeAs
single crystal at different temperatures. $R(\omega)$ in normal
state above $T_c$ (=17.6 K) decreases to follow the Hagen-Rubens
relation in the frequency range from 20 to $\sim$100 cm$^{-1}$,
which indicates the metallic behavior of LiFeAs.  The 4 K data
shows a clear signature of the development of the SC state in
reflectivity when it approaches unity below 25 cm$^{-1}$ within
the uncertainty level of 0.6 \%. This behavior in reflectivity is
an undeniable evidence of the SC gap formation. The bottom inset
of Figure 1 shows the reflectivity spectra for overall range of
the measured frequencies. In the infrared region, two knee-like
steps were observed at 800 and 2,500 cm$^{-1}$. These steps in the
reflectivity spectra are caused by interband transitions. The top
inset of Fig. 1 is the resistivity data, measured by the standard
four probe method, showing the SC transition at 17.6 K with RRR
$\sim$ 22.

\begin{figure}[tbp] \centering
\includegraphics[width=1 \linewidth]{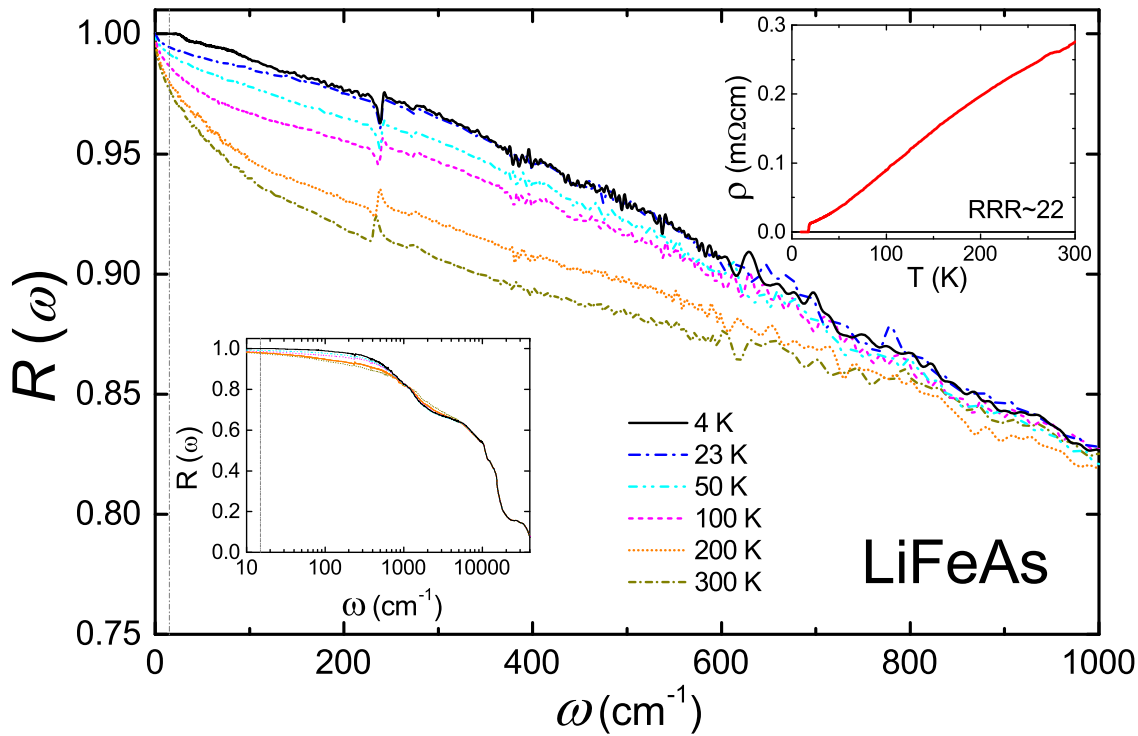} \caption{(color online).
Reflectivity spectra, R($\omega$), of LiFeAs ($T_c$ = 17.6 K) in
the low frequency region for various temperatures. The normal state data below
15 cm$^{-1}$ (dashed line) were extrapolated using the Hagen-Rubens
formula. The bottom inset shows reflectivity in the frequency
up to 40,000 cm$^{-1}$. The top inset shows the electrical
resistivity of LiFeAs single crystal and indicates
the Residual Resistivity Ratio (RRR) $\sim$ 22 of our sample.}\label{figure1}
\end{figure}

For more convenient analysis, the real part of optical
conductivity $\sigma_1(\omega)$ was calculated using the KK
transformation from our reflectivity data. Following the standard
procedure, Hagen-Rubens formula was used for the low frequency
extension below 15 cm$^{-1}$ with the value obtained from
electrical resistivity for normal state. Figure 2 shows the
results of $\sigma_1(\omega)$ at different temperatures. In normal
state, $\sigma_1(\omega)$ decreases from dc value with increasing
frequency which is a typical feature of the Drude response.
Furthermore, the width of the low frequency Drude part of
$\sigma_1(\omega)$ rapidly decreases with decreasing temperature
which indicates the systematic evolution of the coherent metallic
state with decreasing temperature up to $T_c$.

In the mid-IR region, two sharp peaks are observed at 240 and 270
cm$^{-1}$. Jishi $et$ $al$. \cite{jishi} reported the calculated
frequencies of IR-active phonon modes at 228 ($E_{u}$), 276
($E_{u}$), 277 ($E_{u}$) and 338 ($A_{2u}$) cm$^{-1}$ in LiFeAs.
By comparison, these two peaks of our experimental data correspond
to the IR-active phonons. Interesting behavior of these IR-active
phonon peaks is their strong temperature dependence; its peak
intensity becomes rapidly sharper as decreasing temperature.
Similar behavior was observed in infrared study of BaFe$_2$As$_2$
and was explained by orbital ordering in the Fe-As layers
\cite{akrap}. We suspect that a similar orbital ordering might
occur in LiFeAs.

\begin{figure}[tbp] \centering
\includegraphics[width=1 \linewidth]{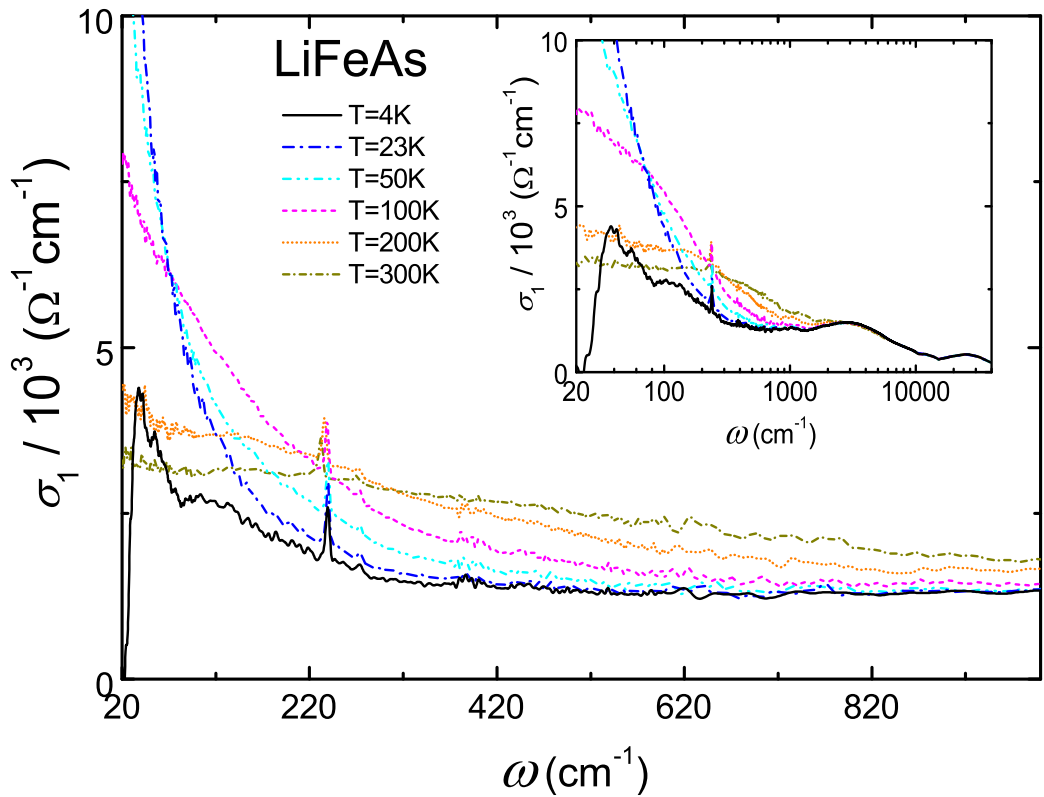} \caption{(color online).
Real part of the optical conductivity, $\sigma_1(\omega)$, of LiFeAs in the
low frequency region for various temperatures. The inset shows  $\sigma_1(\omega)$
up to 40,000 cm$^{-1}$.}\label{figure2}
\end{figure}

Below $T_c$, the 4 K data shows a dramatic change in the low
frequency region: a sudden drop and vanishing of the optical
absorptions below ~25 cm$^{-1}$. This change in the optical
conductivity below $T_c$ should arise from the formation of SC
energy gap. Our 4 K data of $\sigma_1(\omega)$ is practically zero
below ~25 cm$^{-1}$, within the uncertainty level of 0.6 \%. This
complete suppression of the optical absorption is also reflected
in the reflectivity data with $R(\omega) \rightarrow$  1 below
$T_c$ (see Fig.1). In the clean limit superconductivity, no
optical excitations exist at the frequencies lower than twice the
SC gap magnitude (2$\Delta$) \cite{tinkham1}, hence we conclude
that our LiFeAs sample is a clean limit superconductor. Assuming
the sign-changing multiple s-wave pairing state, as generally
accepted for most of the iron based superconductors, this clean
limit opening of the optical absorption gap implies that the
interband impurity scattering is absent or very weak
\cite{Bang_imp}. On the other hand, the fat Drude spectra at
normal state (the full width half maximum of it at 23 K is about
60 cm$^{-1}$) and the significant absorption spectra above the
absorption edge $\omega
> 2\Delta$ in the SC state imply that there should exist a
sufficient amount of scattering both in normal and SC states. The
reconciliation between the clean limit SC behavior and the large
scattering rate even below $T_c$ leads us to the following
scenario for the scattering: (1) the impurity scattering should be
very weak; (2) the strong inelastic scattering exists and its low
energy part is cut off when the system enters the SC phase,
indicating its dynamic coupling with the free carriers of the
Drude component.

\subsection{Drude-Lorentz model analysis}

\begin{figure}[tbp]
\includegraphics[width=1 \linewidth]{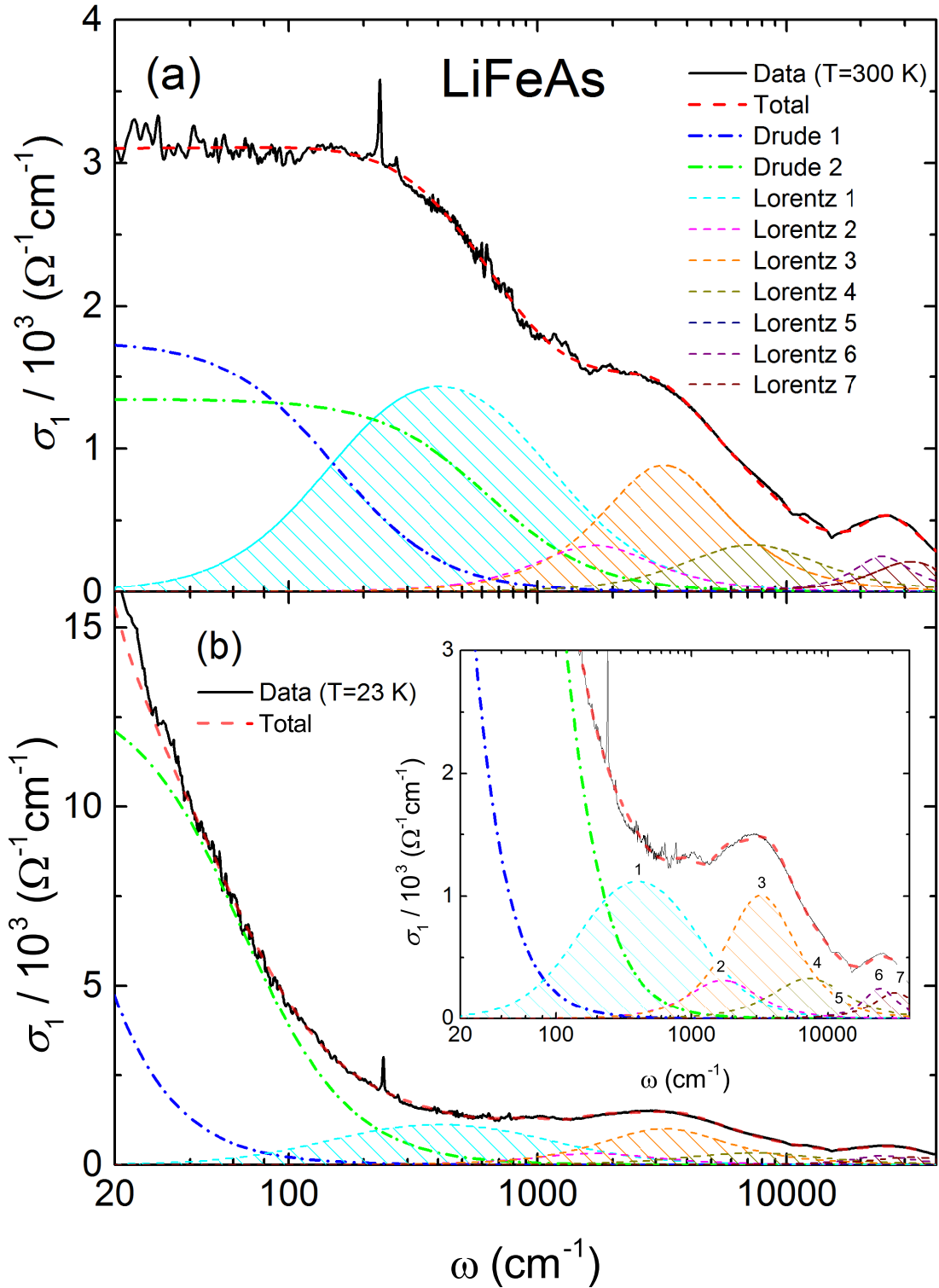}
\caption{(color online). The results of the best fit for the optical
conductivity of the 300 K (a) and 23 K data (b) with two Drude
components and seven Lorentzian oscillators.  Inset of (b) shows a close
up view of the $T$ = 23 K data fitting}\label{figure3}
\end{figure}

In order to understand the further details of the electronic
structure of LiFeAs, we have analyzed the normal state
$\sigma_1(\omega)$ using the standard Drude-Lorentz model in which
the optical absorptions are described by separate contributions of
the delocalized carriers at low frequencies and the excitations of
the bound electrons at the high frequency region. Thus we fit our
data $\sigma_1(\omega)$ using the following formula:
\begin{equation}\label{1}
 \sigma(\omega) = \frac{1}{4\pi}\Bigg[\sum_j
  \frac{\omega^2_{P,j}}{\frac{1}{\tau_{D,j}}-i\omega}
+ \sum_k S_k
  \frac{\omega}{\frac{\omega}{\tau_{L,k}}+i(\omega^2_{0,k}-\omega^2)}\Bigg]
\end{equation}
\noindent where $\omega^2_{P,j}=4\pi n_je^2/m^*_j$ and
$1/\tau_{D,j}$ are the plasma frequency squared, scattering rate
for the $j$-th band, respectively, and $S_k$, $\omega_{0,k}$ and
$1/\tau_{L,k}$ are the strength, center and width of the $k$-th
oscillation, respectively.
First, we tried one Drude band fitting for the low frequency Drude
part of the 23K data but failed, and we found at least two Drude
bands are necessary and the fitting was successful as shown in
Fig.3(b). Then the rest of the high frequency spectral density can
be optimally fitted with seven Lorentzian oscillators. This result
is shown in Fig.3(b) and the fitting parameter values are listed
in Table 1.

\begin{table}
\centering
\begin{tabular}{c}
(A)~Drude Spectra Parameters\\
\end{tabular}

\begin{tabular}{cccc}
\hline\hline
 T (K)  & Drude band & $\omega_{P,j}$ (cm$^{-1}$)  &  1 /$\tau_{D,j}$ (cm$^{-1}$) \\
\hline

   23     &  Drude 1 & 4,033 & 8 \\
            &  Drude 2 & 7,173 & 65 \\
\hline

 300    &  Drude 1 & 4,032 & 154 \\
            &  Drude 2 & 7,173 & 637 \\
\hline             \\
\end{tabular}

\begin{tabular}{c}
(B)~Lorentz Oscillators Parameters\\
\end{tabular}

\begin{tabular}{cccc}
\hline\hline
23K     &  $S_k$ & $~~~~~~\omega_{0,k}$ (cm$^{-1}$) & 1/$\tau_{L,k}$ (cm$^{-1}$) \\
 \hline

 Lorentz 1 & 1.25       & 403   & 1210  \\
 Lorentz 2 & 0.75        & 1694  & 2500  \\
 Lorentz 3 & 4.64      & 3226  & 5001  \\
 Lorentz 4 & 3.71          & 7259  & 12099 \\
 Lorentz 5 & 0.03          & 12099 & 8066  \\
 Lorentz 6 & 3.85          & 24359 & 16535  \\
 Lorentz 7 & 5.60            & 31456 & 28230  \\
 \hline
300K     &  $S_k$ & $~~\omega_{0,k}$ (cm$^{-1}$) & 1/$\tau_{L,k}$ (cm$^{-1}$) \\
 \hline

 Lorentz 1 & 1.60     & 403   & 1210  \\
 Lorentz 2 & 0.75     & 1694  & 2500  \\
 Lorentz 3 & 4.08     & 3226  & 5001  \\
 Lorentz 4 & 3.71         & 7259  & 12099 \\
 Lorentz 5 & 0.03         & 12099 & 8066  \\
 Lorentz 6 & 3.85         & 24359 & 16535  \\
 Lorentz 7 & 5.60           & 31456 & 28230  \\
 \hline
\end{tabular}
\caption{\label{tab:table1} Parameters of the Drude-Lorentz fit of
the optical conductivity of the 23 and 300 K data (Data of Fig.3).
(a) $\omega_{p,j}$ and $1/\tau_{D,j}$ are the plasma frequency and
scattering rate of the $j$-th Drude band, respectively. (b) $S_k$,
$\omega_{0,k}$, and $1/\tau_{L,k}$ are the oscillator strength,
the resonance frequency, and the width of the $k$-th Lorentzian
oscillator, respectively.}
\end{table}

Then we fit the low frequency Drude part of the 300K data with the
same two Drude bands as used in the 23K data. We found that two
Drude bands with the same plasma frequencies but only with the
increased scattering rates fit the data very well. Also the rest
of the high frequency spectra was well fitted with the same seven
Lorentzian oscillators with almost same fitting parameters as used
in the 23K data. This result is summarized in Table 1. Therefore,
the main difference between the 23K data and 300K data is the
temperature evolution of the Drude part with decreasing scattering
rates with decreasing temperature. On the other hand, the
Lorentzian oscillator part shows almost no change with temperature
and we believe that their origin is the interband transition as
confirmed with the band calculations in the next section.
The total plasma frequency estimated by $f$-sum rule of the two
fitted Drude terms is $\omega_p$ = 8224.4 cm$^{-1}$, which is
about 10 \% smaller than that of 122 superconductors reported
already \cite{song2,kimh}. The scattering rates of the Drude
spectra are also smaller compared to the other Fe-based
superconductors\cite{song2,kimh}. These facts might be
concomitantly related with the moderate $T_c \approx 17K$ of
LiFeAs. To have more comparison, the optimal doped 11 compound,
FeTe$_{0.55}$Se$_{0.45}$ \cite{homes}, which has slightly lower
$T_c \sim 14$K than our 111 compound, has a slightly smaller value
of the total Drude plasma frequency $\omega_p \sim$ 7200
cm$^{-1}$.

We can also extract some more information from our fitting values
of Drude spectra. The plasma frequency of Drude-2 band is much
larger than that of Drude-1 band, which indicates that the FS of
the Drude-2 band is much larger than the FS of the Drude-1 band.
Also, the drastic decrease of the scattering rates
($1/\tau_{D,j}$) from 300K to 23K (see Table 1.(A)) indicates that
the dominant scattering process must be of inelastic origin and
the contribution from the impurity scattering must be very weak.
This is also consistent with the fact that the most possible
origin of impurities in LiFeAs is the Li vacancies which are
located above the conducting Fe-As layers. The analysis using the
Drude-Lorentz model with two Drude components was already employed
in several optical spectroscopy studies of the iron based
superconductors \cite{kimk, tu, akrap}. Indeed, various other
experimental and theoretical studies also pointed out the
multiband features and the weak interband scattering in LiFeAs
superconductor \cite{hajiri, ferber, hashimoto, umezawa}.

\subsection{Lorentz oscillators and Interband transitions}

In order to have a direct comparison of the theoretical electronic
structure of LiFeAs with our optical measurement, we have
calculated the direct interband transitions in $\sigma_1(\omega)$
using the band calculation results and have compared them to the
Lorentz oscillators of our optical conductivity data. The LDA
calculation for the band structures was performed with WIEN2k code
and the interband transitions of $\sigma_1(\omega)$ spectra were
derived from a standard formula as follows \cite{antonov};

\begin{figure}[tbp]\centering
\includegraphics[width=1 \linewidth]{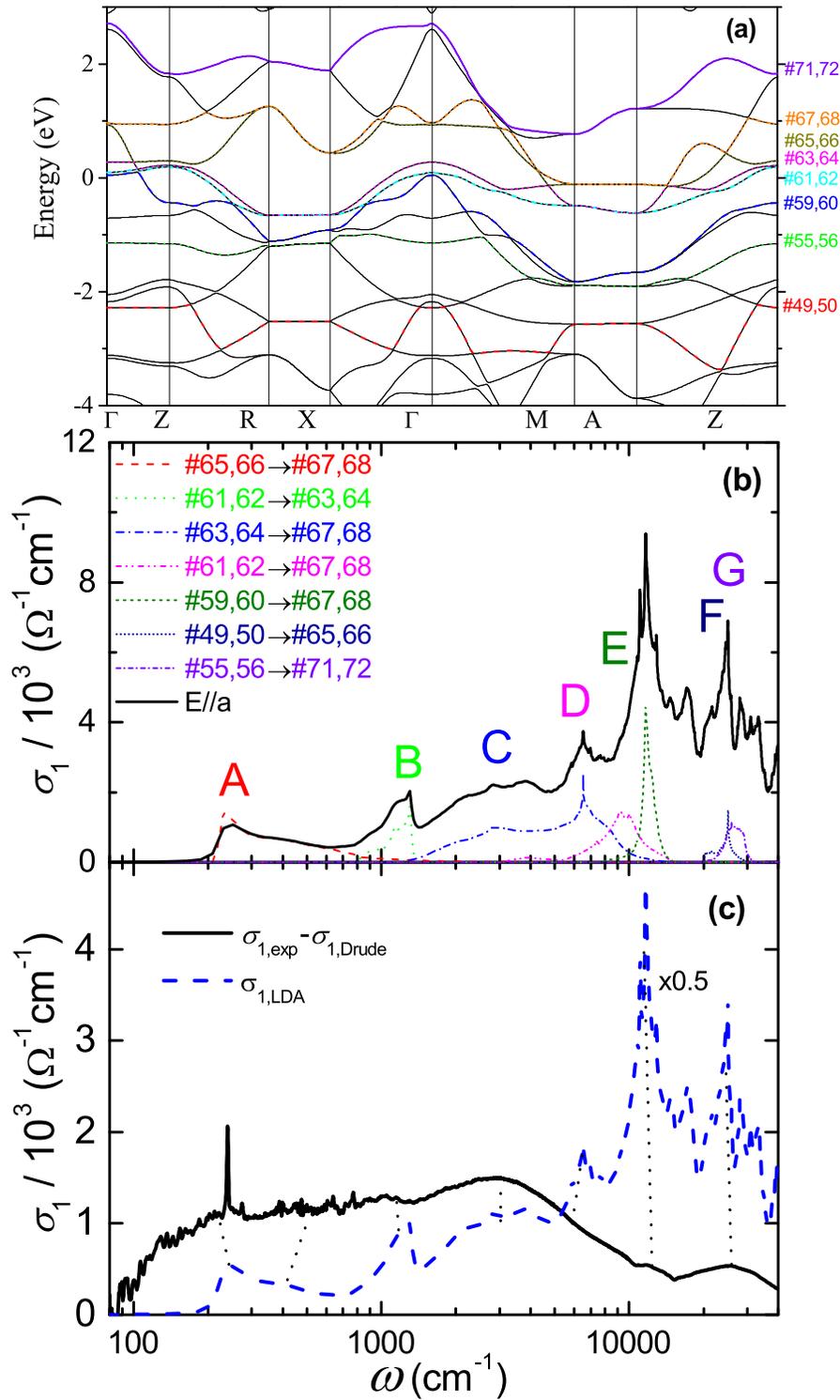} \caption{(color online).
(a) Calculated band structure of LiFeAs near the Fermi level, (b)
Calculated interband transition contributions to $\sigma_1(\omega)$ of LiFeAs.
The total (black solid) and the separated band-to-band contributions.
(A: \#65,66 $\rightarrow$ \#67.68, B: \#61,62$\rightarrow$ \#63.64,
C: \#63,64 $\rightarrow$ \#67.68, D: \#61,62 $\rightarrow$ \#67.68,
E: \#59,60 $\rightarrow$ \#67,68, F: \#49,50 $\rightarrow$ \#65,66 and
G: \#55,56 $\rightarrow$ \#71,72),
(c) Comparison between the calculated (blue dashed line) and
experimental (black solid line) optical conductivity spectra at $T$ =23 K. }\label{figure4}
\end{figure}

\begin{equation}
\sigma(\omega) = \frac{\pi e^2}{m^2_0\omega}\sum_{\bf{k}}\sum_{nn'}
\frac{|<n' \bf{k}|\bf{e}\cdot\bf{p}|\emph{n}
\bf{k}>|^2}{\omega-\omega_{nn'}(\bf{k})+\emph{i}\bf{\Gamma}}
\times
\frac{f(\epsilon_{n\bf{k}})-f(\epsilon_{n'\bf{k}})}{\omega_{nn'}(\bf{k})}
\nonumber
\end{equation}

\noindent Here, the $|n' \bf{k}>$ and $|n \bf{k}>$ states denote
the unoccupied and occupied states, respectively, {\bf e} and {\bf
p} are the polarization of light and the momentum of the
electron,respectively, $f(\varepsilon_{n\bf{k}})$ is the
Fermi-Dirac distribution function, $\hbar\omega_{nm} =
\varepsilon_{n\bf{k}} - \varepsilon_{m\bf{k}}$ is the energy
difference between the unoccupied and occupied states, and
$\Gamma$ is the lifetime. In the calculation, $\Gamma$ = 1 meV was
assumed. The band structure near the Fermi level is shown in Fig.
4 (a) with some of the bands labelled. The experimental result at
$T$ = 23 K and the calculated spectra of $\sigma_1(\omega)$ are
displayed together in Figure 4 (b) and 4 (c). Here, two Drude
parts from the fit were subtracted from the experimental data
because the calculation with Eq. (2) included only the interband
transition processes. The calculated optical spectra have peaks at
around 250, 1,300, 3,000, 6,500, 12,000 and 25,000 cm$^{-1}$,
respectively. The origin of each peak is denoted in the legend of
Figure 4 (b) as A to G. For example, the peak at 1,300 cm$^{-1}$
(denoted as B) is due to the interband transitions from \#61, 62
to \#63, 64 labelled bands. The peak positions are in good
agreement with experimental result (pointed by dotted lines in
Figure 4 (b)) while the peak intensities are not as good in
agreement with the experimental data as in the peak position. This
is understandable because Eq. (2) is using a very simple coupling
matrix $\sim 1/m_0$ and the actual optical coupling matrices
should be more complicated. The overall intensity of the
calculated $\sigma_1(\omega)$ below 10,000 cm$^{-1}$ is
qualitatively consistent with experimental spectra. However, the
intensity of the calculated $\sigma_1(\omega)$ above 10,000
cm$^{-1}$ is much larger than the experimental value. Also the
spectra above 10,000 cm$^{-1}$ have large overlap of the multitude
of the transitions between several bands thus the origin of the
peaks becomes harder to identify.

\subsection{Optical Conductivity in Superconducting State}

As shown in Fig. 2, the change of the optical conductivity from 23
K to 4 K in the low frequency range clearly indicates the
formation of a SC energy gap. The SC plasma frequency
($\omega_{ps}$) can be estimated from the spectral weight
transfer. According to the Ferrell-Glover-Tinkham (FGT) sum rule
\cite{ferrell, tinkham2}, the spectral weight difference in the
optical conductivity data between just above $T_c$ and much below
$T_c$ (so called the missing area) indicates the condensation
strength and determines the condensation density of the free
carriers, which is described as follows \cite{ferrell, tinkham2}:

\begin{equation}
\omega^2_{ps} = 8\int^{\omega_c}_0\Big[ \sigma_1(\omega, T \cong T_c)
- \sigma_1(\omega, T \ll T_c)\Big]d\omega
\end{equation}

\noindent where $\omega^2_{ps}$ = 4$\pi n_s e^2/m^*$. The cut-off
frequency ($\omega_c$) was set as 1,000 cm$^{-1}$ because there is
almost no difference between 23 K and 4 K data. SC plasma
frequency ($\omega_{ps}$) can also be evaluated by the zero
crossing of the dielectric function, $\epsilon_1(\omega)\approx
\epsilon_{\infty} - \omega^2_{ps}/\omega^2$ ($\epsilon_{\infty}
\approx 3.6$), and by the zero frequency limit of the real part of
$[-\omega^2\varepsilon_1(\omega)]^{0.5}$. All three methods
consistently yield $\omega_{ps}$  $\sim$ 6,665 $\pm$ 140 cm$^{-1}$
. Combining with the previous estimate of $\omega_p$ = 8224.4
cm$^{-1}$, we obtained ($\omega_{ps}/\omega_p)^2$ $\sim$ 0.65,
which indicates that more than a half of the free carriers of the
normal state condensates.
The penetration depth evaluated from the relation $\lambda =
c/\omega_{ps}$  is 238 nm, which is about 10\% larger than the
already reported results \cite{kimh,song2,pratt}.

\begin{figure}[tbp]\centering
\includegraphics[width=1 \linewidth]{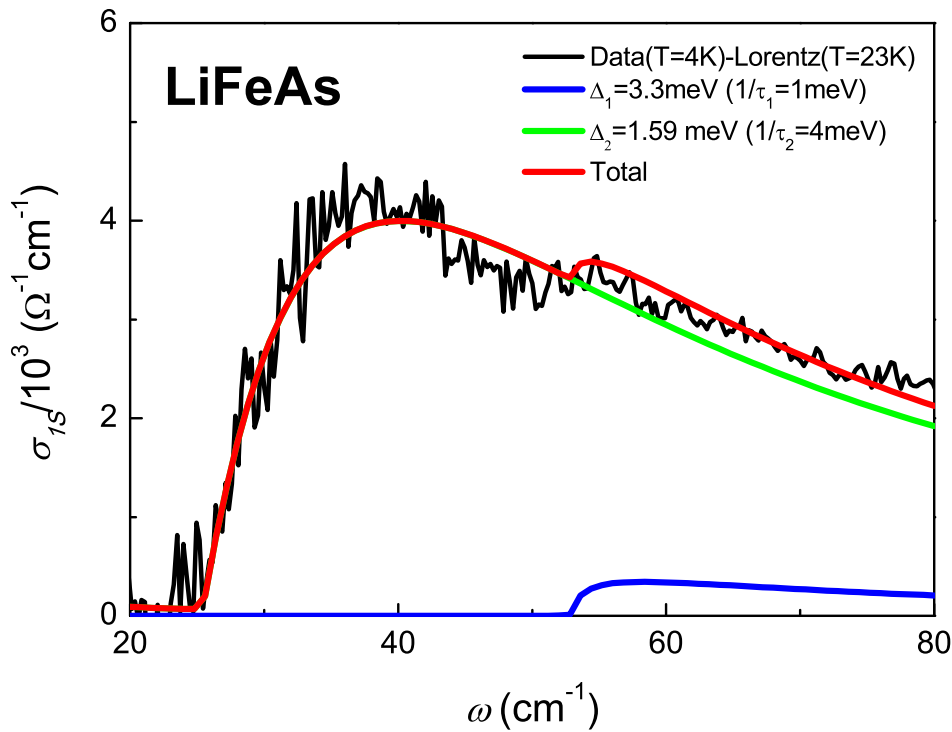} \caption{(color online).
Fitting of the low frequency optical conductivity ($\sigma_{1S}$)
of SC state (4 K data) using the generalized Mattis-Bardeen formula
\cite{zimmermann}. Two bands fitting is required and the estimated SC
energy gap values, corresponding to the Drude-1
and Drude-2 bands of the normal state,
are $\Delta_1 =3.3$ meV and $\Delta_2 =1.59$ meV with
the corresponding scattering rates $1/\tau_1 =1$ meV and $1/\tau_2 =4$ meV,
respectively.}\label{figure5}
\end{figure}

The low frequency optical conductivity ($\sigma_{1S}$) of the 4K
data is separately displayed in Figure 5. The data clearly shows
the opening of the optical absorption gap below approximately 25
cm$^{-1}$ due to the formation of the SC gap. The theoretical
calculation based on the isotropic s-wave gaps using the
generalized Mattis-Bardeen formula \cite{zimmermann} was used to
fit the data. Here we fitted the data of $\sigma_{1S}(\omega)$
with two bands with two independent s-wave gaps, which is
consistent with the two Drude components analysis of the normal
state $\sigma_{1N}(\omega)$ in the previous section. The results
are in excellent agreement with the experimental data as shown in
Figure 5. The SC gaps (and scattering rates) were estimated to be
$\Delta_1$ = 3.3 meV ($1/\tau_1$ = 1 meV) and $\Delta_2$ = 1.59
meV ($1/\tau_2 = 4$ meV). The gap to $T_c$ ratios are,
$2\Delta_{1,2}/k_B T_c \sim$ 4.5 and 2.17, respectively, as
compared to the BCS weak coupling limit (= 3.5). These values may
be consistently compared with other iron-based SC compounds; for
example, the optimal doped 11-compound, FeTe$_{0.55}$Se$_{0.45}$
($T_c \sim 14$K)\cite{homes}, has slightly smaller values of the
SC gaps, $\Delta_1$ = 2.5 meV and $\Delta_2$ = 1.25 meV,
respectively.

We also found that the band with a larger spectral weight (Drude-2
band, $\omega_{P,2}=7,173$ cm$^{-1}$) opens a smaller gap
$\Delta_2$= 1.59 meV and the band with a smaller spectral weight
(Drude-1 band, $\omega_{P,2}=4,033$ cm$^{-1}$) opens a larger gap
$\Delta_1$= 3.3 meV. This inverse proportionality between the SC
gap size and the spectral weight of the two Drude bands is
consistent with the prediction of the s$_{\pm}$-wave pairing
scenario mediated by the interband repulsive
interaction\cite{Bang_pairing}. Our result of the two SC gaps is
consistent with the observation of other experiments by specific
heat \cite{Stockert}, ARPES \cite{umezawa}, NMR\cite{zhengli}
measurements, and also with a theoretical
prediction\cite{christian}.
Also the scattering rates $1/\tau_{1,2}$ obtained from the
Mattis-Bardeen formula are consistently close to the values of
$1/\tau_{D,j}$ of the Drude-1 and the Drude-2 bands of the normal
state, indicating that the SC gaps are indeed formed in the
Drude-1 and the Drude-2 bands, respectively. Some discrepancy is
due to the fact that the generalized Mattis-Bardeen formula is not
directly derived from the Drude formula.

\section{Conclusions}

We have measured the optical properties of the iron based
superconductor LiFeAs single crystal ($T_c$ = 17.6 K) at various
temperatures from the tera-hertz to violet frequency regions and
have successfully -- for the first time with the optical
spectroscopy - deduced the multi-band nature of LiFeAs both in the
SC and normal states. The optical spectra in the normal state is
well described by the Drude-Lorentz model assuming two Drude
components with $\omega_{D,1}$ = 4,033 cm$^{-1}$ and
$\omega_{D,2}$ = 7,173 cm$^{-1}$, respectively. In the SC state at
T = 4 K, a clean gap opening is observed in our optical
conductivity data below $T_c$ and the theoretical fitting using
the generalized Mattis-Bardeen model \cite{zimmermann} identifies
the two isotropic SC gaps of $\Delta_1$ = 3.3 meV and $\Delta_2$ =
1.59 meV, respectively. These results confirm that the multi-band
nature is essential to understand the electronic properties of
LiFeAs both in the normal state and SC state in accord with
various other experiments. Furthermore, we have extracted the
inverse proportionality between the SC gap size and the spectral
weight of the two Drude bands. This finding is an indirect
evidence supporting the pairing scenario mediated by the interband
pairing interaction\cite{Bang_pairing}.

The total SC plasma frequency was estimated $\omega_{ps} \sim$
6,665 cm$^{-1}$ and it corresponds to an effective penetration
depth of $\lambda$ = 238 nm. From the comparison with the total
normal state plasma frequency $\omega_{p} \sim$ 8,224.4 cm$^{-1}$,
this implies that about 65 \% of the free carriers of the normal
state condenses in the SC state and about 35\% of the free
carriers still remains un-condensed. As seen in Figure 5, the
existence of the substantial amount of the un-condensed incoherent
spectra above the optical gap as well as the clean gap opening
below $\sim$25 cm$^{-1}$ reveal the several important facts: (1)
our LiFeAs single crystal is a clean limit superconductor with a
very weak impurity scattering; (2) nevertheless, it has a strong
inelastic scattering which causes the pair-breaking process above
the optical gap; (3) and this inelastic scattering should also
develop the excitation gap as a low energy cut-off when the system
enters the SC state, meaning that this bosonic inelastic
scattering is dynamically coupled to the free carriers.
Finally, we identified the several Lorentzian oscillators observed
in our optical data over the mid-IR to violet region with the
interband optical transitions by using the LDA band structure
calculations.

\ack{This work was supported by the Basic Science Research Program
(2010-0007487), the Mid-career Researcher Program (2010-0029136),
and the Leading Foreign Research Institute Recruitment Program
(Grant No. 2012K1A4A3053565) through the National Research
Foundation of Korea (NRF) funded by the Ministry of Education,
Science and Technology (MEST). YB was supported by grant numbers
NRF-2010-0009523 and NRF-2011-0017079.}

\end{document}